\documentclass[]{spie}

\usepackage{amsmath,amsfonts,amssymb}
\usepackage{graphicx}
\usepackage[colorlinks=true, allcolors=blue]{hyperref}

\title{Design and testing of Kinetic Inductance Detector package for the Terahertz Intensity Mapper}

\author[a]{L.-J. Liu}
\author[a,b]{R.M.J Janssen}
\author[a,b]{C.M. Bradford}
\author[a]{S. Hailey-Dunsheath}
\author[c]{J.P. Filippini}
\author[d]{J.E. Aguirre}
\author[d]{J.S. Bracks}
\author[d]{A.J. Corso}
\author[c]{J. Fu}
\author[e]{C. Groppi}
\author[e]{J. Hoh}
\author[f]{R.P. Keenan}
\author[f]{I.N. Lowe}
\author[f]{D.P. Marrone}
\author[e]{P. Mauskopf}
\author[c]{R. Nie}
\author[a]{J. Redford}
\author[g]{I. Trumper}
\author[c]{J.D. Vieira}
\affil[a]{California Institute of Technology, Pasadena, CA 91125, USA}
\affil[b]{NASA Jet Propulsion Laboratory, Pasadena, CA 91109, USA}
\affil[c]{University of Illinois at Urbana-Champaign, Urbana, IL 61820, USA}
\affil[d]{University of Pennsylvania, Philadelphia, PA 19104, USA}
\affil[e]{Arizona State University, Tempe, AZ 85281, USA}
\affil[f]{University of Arizona, Tucson, AZ 85721, USA}
\affil[g]{Intuitive Optical Design Lab LLC, Tucson, AZ 85701, USA}

\authorinfo{Send correspondence to Lun-Jun (Simon) Liu: E-mail: lliu@caltech.edu}

\begin{document} 
\maketitle

\begin{abstract}
The Terahertz Intensity Mapper (TIM) is designed to probe the star formation history in dust-obscured star-forming galaxies around the peak of cosmic star formation. This will be done via measurements of the redshifted 157.7 $\mathrm{\mu m}$ line of singly ionized carbon ([CII]). TIM employs two R $\sim 250$ long-slit grating spectrometers covering $240-420$ $\mathrm{\mu m}$. Each is equipped with a focal plane unit containing 4 wafer-sized subarrays of horn-coupled aluminum kinetic inductance detectors (KIDs). We present the design and performance of a prototype focal plane assembly for one of TIM's KID-based subarrays. The overall detector package must satisfy thermal and mechanical requirements, while maintaining high optical efficiency and a suitable electromagnetic environment for the KIDs. In particular, our design manages to strictly maintain a 50 $\mathrm{\mu m}$ airgap between the array and the horn block. The prototype detector housing in combination with the first flight-like quadrant are tested at 250 mK. A frequency scan using a vector network analyzer shows 823 resonance features, which represents $>90\%$ yield, indicating a good performance of our TIM detector wafer and the whole focal plane unit. Initial measurements also show that many resonances are affected by collisions and/or very shallow transmission dips as a result of a degraded internal quality factor (Q factor). This is attributed to the presence of an external magnetic field during cooldown. We report on a study of magnetic field dependence of the Q factor of our quadrant array. We implement a Helmholtz coil to vary the magnetic field at the detectors by (partially) nulling earth's. Our investigation shows that the earth magnetic field can significantly affect our KIDs' performance by degrading the Q factor by a factor of $2-5$, well below those expected from the operational temperature or optical loading. We find that we can sufficiently recover our detectors' quality factor by tuning the current in the coils to generate a field that matches earth's magnetic field in magnitude to within a few $\mathrm{\mu T}$. We emphasize that it is impractical to fly a Helmholtz coil on TIM and dynamically ``null'' earth's field. Therefore, it is necessary to employ a properly designed magnetic shield enclosing the TIM focal plane unit. Based on the results presented in this paper, we set a shielding requirement of $|B| <$ 3 $\mathrm{\mu T}$.

\end{abstract}

\keywords{Astronomy, Terahertz, Far-infrared, Balloon, Spectroscopy, Kinetic inductance detector, Aluminum, Line intensity mapping}

\section{INTRODUCTION}
\label{sec:intro}

Understanding the formation and evolution of stars and galaxies over cosmic history is one of the foremost goals of astrophysics and cosmology~\cite{JV2019}. The cosmic star formation rate (SFR) has undergone a dramatic evolution over the course of 14 billion years, where total SFR peaks around redshift 2 and then declines rapidly towards the current Universe~\cite{PM2014}. Luminous infrared (IR) galaxies play a crucial role in the evolution of the cosmic SFR as they contribute more than half of all star formation in the Universe. However, these sources are rich in dust, which absorbs the radiation of stars and re-radiates it in IR wavelengths~\cite{MH2001,GL2005}. The heavy dust extinction makes it challenging to probe the star formation activity in these sources using conventional (optical) surveys, but the re-radiation in IR implies that far-IR spectroscopy is a vital tool to discover a wealth of information dust-obscured galaxies can provide on the nature of star formation and galaxy evolution.

We are building the Terahertz Intensity Mapper (TIM), a balloon-borne experiment with the goal of unravelling the formation and evolution of galaxies around the peak of cosmic star formation ($0.5<z<1.7$). TIM will perform spectroscopic measurements of redshifted far-IR spectral lines, most importantly the [CII] 157.7 $\mathrm{\mu m}$ line, to provide an unbiased dust-immune census of the star formation history~\cite{BU2014}. Applying a novel methodology called line intensity mapping enables TIM to measure the total line intensity of all galaxies in the volume set by its spectral and spatial resolution, which can thus determine the three-dimensional structure of our star-forming Universe.

Successful deployment of TIM will not only advance our understanding of galaxy evolution through observations that cannot be accomplished by current far-IR instruments, TIM will also be a vital technological stepping stone for future space missions. TIM employs two R $\sim 250$ long-slit grating spectrometers covering $240-420$ $\mathrm{\mu m}$, consisting of a short wavelength (SW) band covering $240-317$ $\mathrm{\mu m}$ and a long wavelength (LW) band covering $317-420$ $\mathrm{\mu m}$. Each band is equipped with a focal plane unit containing 4 wafer-sized subarrays (termed quadrants) of $\sim 900$ Kinetic Inductance Detectors (KIDs). KIDs are a key technological development for future far-IR space missions, and the KIDs on TIM will provide the closest analog in terms of sensitivity and operational environment.

In this paper we present the design of single quadrant detector package and the study on its mechanical and electromagnetic shielding properties. In particular, we investigate the magnetic field dependence of the KIDs designed for use in LW spectrometer of TIM. In Sec.~\ref{sec:design} we will describe the design of the subarray focal plane unit, which will be a prototype of the larger TIM flight focal planes. In Sec.~\ref{sec:testing} we will show the high yield of detectors in this housing through a full frequency scan using a vector network analyzer, and we will report on the performance of the detectors in various magnetic fields. These shall demonstrate the robustness of our package design and the necessity of implementing a magnetic shield. A summary of the key properties of package design and the magnetic field study is given in Sec.~\ref{sec:summary}.

\section{DESIGN OF QUADRANT FOCAL PLANE ASSEMBLY}
\label{sec:design}

\begin{figure} [ht]
   \begin{center}
   \begin{tabular}{c}
   \includegraphics[height=6cm]{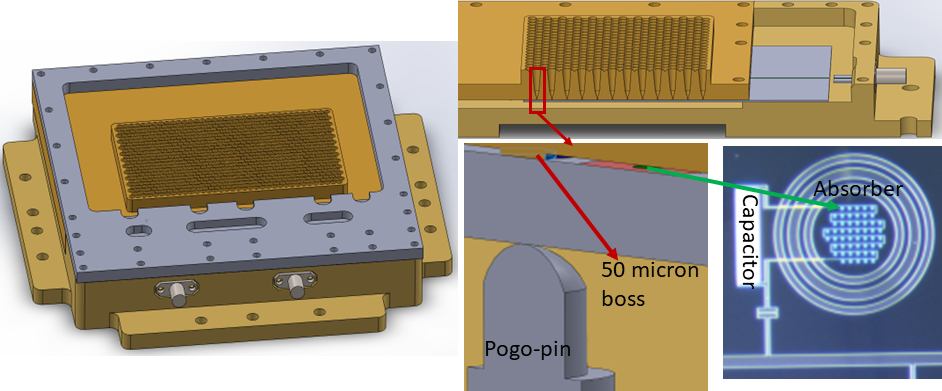}
   \end{tabular}
   \end{center}
   \caption[quad] 
   { \label{fig:quad} 
Left: The solid model of the LW quadrant detector package. Right: A zoom of the wafer mount architecture. The 50 micron bosses are pressed onto the front of the detector wafer by a matching array of spring-loaded pins pushing against the wafer from the bottom. Bottom right: A photograph of a single KID as seen through a microscope.}
   \end{figure}

For the purpose of testing individual detector wafer, we fabricate a 864 pixel array of flight-like KIDs as a prototype of TIM's LW focal plane. KIDs are thin-film superconducting micro-resonators which absorb incident radiation and respond by changing the resonance frequency and quality factor (Q factor)~\cite{PD2003,JZ2012}. TIM uses a lumped-element KID design made from 30 nm aluminum ($T_c = 1.28$ K) as shown in the bottom right panel of Figure~\ref{fig:quad}. The lumped element KID consist of an inductive absorber and an interdigitated capacitor. The geometry of the latter is tuned to give each of the detectors in this quadrant subarray a unique resonance frequency. As such, we are able to spread all 864 detectors near-uniformly throughout the entire 0.5 GHz readout bandwidth. Each KID is capacitively coupled to the microstrip feed line with the coupling Q factor ($Q_c$) designed to be $10^5$. Such a design enables the quadrant to be read out through a single microstrip readout line using ROACH2 or RFSoC based readout system~\cite{JV2019}. These KIDs have shown photon-noise-limited operation under 100 fW of optical loading at an operational temperature of 250 mK. A detailed description of the KID array design and the single pixel performance can be found in Janssen et al.~\cite{RJ2022}.

The design of individual detector's inductor geometry should secure a high absorption efficiency in the far-IR. We enable this through implementing of a novel ``chain-link'' (CL) design as the unit cell of inductive absorber and then completing the entire absorber via spreading 52 CL unit cells on a single meander. Using ANSYS-HFSS\footnote{https://www.ansys.com/products/electronics/ansys-hfss} simulation, we demonstrate that the CL absorber offers $\sim 90\%$ efficiency in both linear polarization modes. Details of the CL absorber design, simulation and optimization are presented in Nie et al.~\cite{RN2022}.

Each individual KID is optically coupled via a matching array of direct-machined feedhorns in an aluminum horn block. Aiming to ensure high absorption efficiency in this coupling architecture, we carry out a design of our detector package which (1) maintains a small and steady gap between the array chip and the feedhorn block and (2) ensures both the airgap and the detector array survive thermal cycling. Our approach requires a well-controlled 50 micron spacing that also minimizes optical crosstalk. We enable this with an array of spring-loaded pins pushing the chip against a matching array of precision-machined bosses on the bottom of the horn block, one of which is shown in the bottom middle panel of Figure~\ref{fig:quad}. This approach offers compliance to thermal deformations on cooldown, while avoiding any front-side circuitry~\cite{LJL2022}. As a result, our microstrip architecture of the readout line utilizes the horn block surface facing the silicon as the ground.

We have implemented a quadrant-sized detector package to serve as both a prototype of the flight hardware and as a facility to test individual KID arrays before integration into the flight cryostat. This design is a smaller version of the flight-hardware design presented by Liu et al.~\cite{LJL2022}. Figure~\ref{fig:quad} (left) shows the overall structure of our quadrant detector package. From top to bottom, the components are (1) an aluminum cap which presses down the wire bonding section of detector wafer using spring-loaded pins, (2) an aluminum horn block with feedhorns for the quadrant subarray, (3) a quadrant detector wafer containing 864 KIDs, and (4) an aluminum bottom housing which mounts an array of spring-loaded pins for clamping the wafer.

\section{TESTING OF 864 PIXEL LW QUADRANT ARRAY}
\label{sec:testing}

The quadrant-sized detector package has been prototyped and tested at 250 mK, in which the 864 pixel flight-like quadrant is mounted. This initial cooldown of the quadrant subarray is conducted without the mu-metal magnetic shield. A frequency scan within the design frequency range from 360 MHz to 800 MHz shows 823 resonance features, which represents $>90\%$ yield. Such a high yield demonstrates that our electromagnetic design of the TIM detector wafer and the system is realistic, and that the mechanical architecture is robust~\cite{LJL2022}. The initial measurement also shows that many resonances are affected by collisions and/or very shallow transmission troughs as a result of a degraded internal quality factor ($Q_i$). This is attributed to the presence of an external magnetic field during cooldown~\cite{DF2016}. For a balloon payload in suborbital space around the south pole, such as TIM, the dominant contribution to the external magnetic field would be earth's field. Therefore, the following cryogenic tests are conducted to quantify the effect of magnetic fields on our aluminum KID array.

\subsection{Experimental Setup of Magnetic Field Dependence Study}
\begin{figure} [ht]
   \begin{center}
   \begin{tabular}{c}
   \includegraphics[height=7cm]{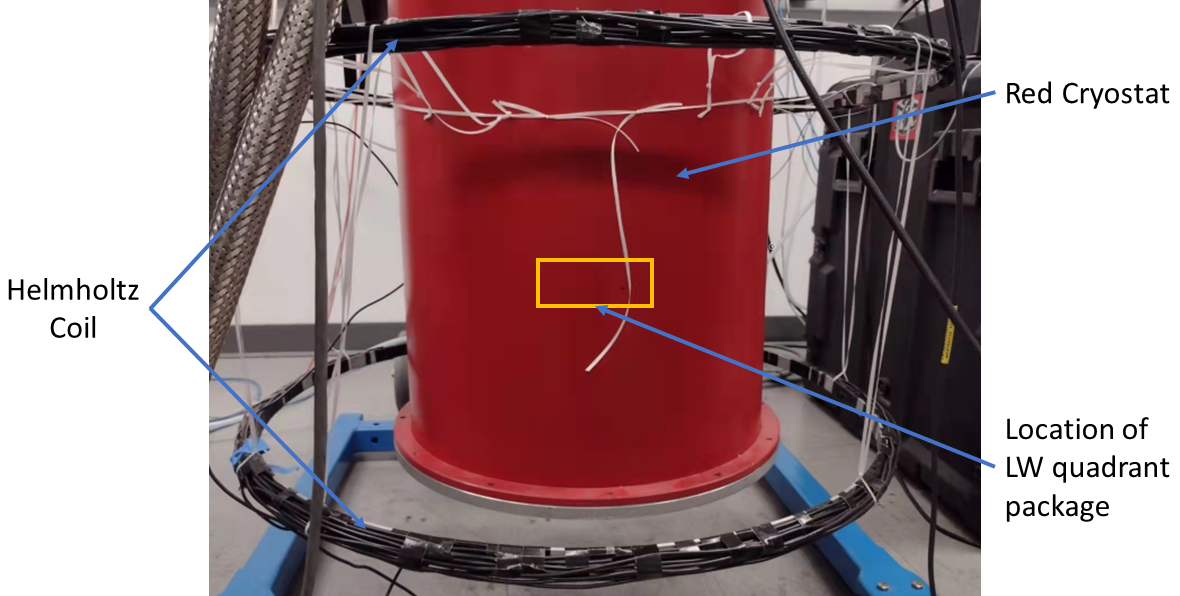}
   \end{tabular}
   \end{center}
   \caption[coil] 
   { \label{fig:coil} 
The experimental setup of Helmholtz coil measurement. The Helmholtz coil is properly hung so that our LW subarray assembly is on the middle plane where the generated magnetic field is uniform.}
   \end{figure}

For these cryogenic tests on our quadrant detector chip, the entire KID array is blocked by a solid aluminum lid, i.e. no optical coupling via horns is present. We have done 2 cooldowns in contrast, one with the cylindrical mu-metal shield enclosing the LW package and the other without. Specifically, in the first cooldown the shield is located inside the cryostat's vaccuum jacket with one side closed (the side near the bottom in Figure~\ref{fig:coil}). The shield diameter is 269 mm and its height equals 219 mm, and the package is located at approximate center of this shield. In order to further quantify the dependence on the earth magnetic field, we employ a Helmholtz coil in the second cooldown without the shield. As displayed in Figure~\ref{fig:coil}, the Helmholtz coil is a pair of identical coils hung concentrically with a vertical separate distance equals the radius of the coil. When properly placing the Helmholtz coil to locate our LW quadrant assembly on the center of the circle and the middle plane between the 2 coils, it generates a uniform magnetic field through the focal plane unit. In such a configuration, we can input desirable current to generate a field opposite to that of the earth's field and thus can null it. Under various nulling fields, we measure the response of KIDs. It is worth mentioning that for every current input (i.e., magnetic field generated by the coil), we must cycle the fridge to lock the nulling field inside the cryostat. This cycling process only heats the experimental stages inside the cryostat, indicating that both the aluminum package and aluminum KIDs temporarily become normal before going through transition and freezing in field, while it is expected that the KIDs go through the transition first as they are made of thin film aluminum and so have higher transition temperature ($T_c$) than that of the bulk aluminum package. The above 2 cooldowns are done via a pulse-tube pre-cooled triple-stage He-10 sorption cooler.

We obtain the resonance frequency and Q factors of all KIDs in the array using a vector network analyzer (VNA). We extract the magnitude plot of $S_{21}$ from the VNA scan, where the overall transmission is normalized around 0 dB for convenience of identifying individual resonance features. We assess the Q factors for each KID's resonance, as high Q values are required for efficient KID multiplexing. When resonances are sufficiently isolated, each resonance line profile is imprinted on the $S_{21}$ plot, where on resonance the $S_{21}$ value reaches the local minimum that can be represented by the following 2 equations,

\begin{equation}
\label{eq:s21min}
| S_{21} |_{min} = 1 - \frac{Q_{r}}{Q_{c}} \, ,
\end{equation}
and

\begin{equation}
\label{eq:totalq}
Q_{r} = Q_{i}^{-1} + Q_{c}^{-1} \, ,
\end{equation}
where $Q_{r}$ is the total Q factor of a KID, $Q_i$ is the internal Q factor which describes the energy dissipated within a resonator, $Q_c$ is the coupling Q factor indicating the loss between the detector and readout feedline. Fitting the magnitude line profile of all identified KIDs, we are able to extract the distribution of $Q_i$ and $Q_c$.

\subsection{Testing Results}
\begin{figure} [ht]
   \begin{center}
   \begin{tabular}{c}
   \includegraphics[height=14cm]{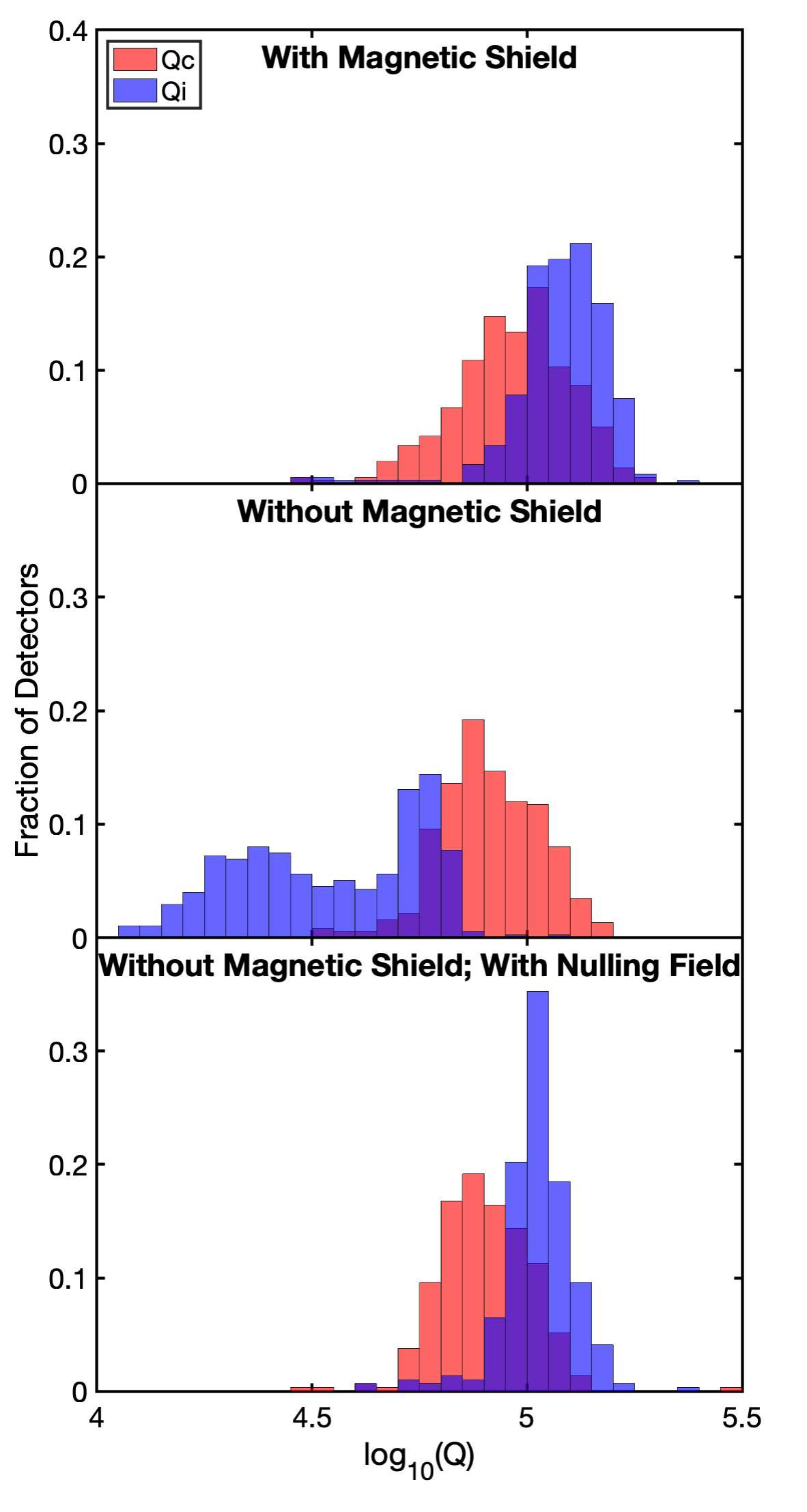}
   \end{tabular}
   \end{center}
   \caption[qfactor] 
   { \label{fig:qfactor} 
Distribution of Q factors for the same KID array with a cylindrical magnetic shield, without magnetic shield, and without magnetic shield but with Helmholtz coil nulling field. For the bottom panel, the current input through the coil is tuned to remove magnetic field as much as possible.}
   \end{figure}

The distribution of $Q_i$ and $Q_c$ in various configurations is presented in Figure~\ref{fig:qfactor}. The top panel shows the statistics of Q factors when the quadrant focal plane assembly is enclosed by the magnetic shield. Distribution of $Q_c$ is centered around $10^{5}$ as designed, while the majority of KIDs has $Q_i > 10^{5}$.

We next remove the shield so that the detector array is exposed under the earth magnetic field. Using 2 smartphone magnetometer apps, we are able measure the strength of the earth's field for reference, which is $\sim 35$ $\mathrm{\mu T}$. The Q distribution extracted from this cooldown is plotted in the middle panel of Figure~\ref{fig:qfactor}. The result indicates that while the distribution of $Q_c$ shows little decrease (from centering around $10^{4.9}$ to $\sim 10^{4.8}$), $Q_i$ decreases by a factor of $2-5$ compared to the measurement with a magnetic shield. More importantly, this reduction would make the resonator $Q_i$ limited by the magnetic field, instead of the optical loading / temperature during operation, from both of which we expect $Q_i \sim 10^{5}$. This Q degradation causes more collisions and puts more stringent requirements on resonator spacing. The distribution of $Q_i$ values also shows a bimodality, with one peak at $\sim 10^{4.75}$ and the other $\sim 10^{4.3}$. This could be due to the fact that in an aluminum detector enclosure, there may be a spatial variation in the trapped magnetic flux, which can degrade the performance of some KIDs more than others.

   \begin{figure} [ht]
   \begin{center}
   \begin{tabular}{c}
   \includegraphics[height=6cm]{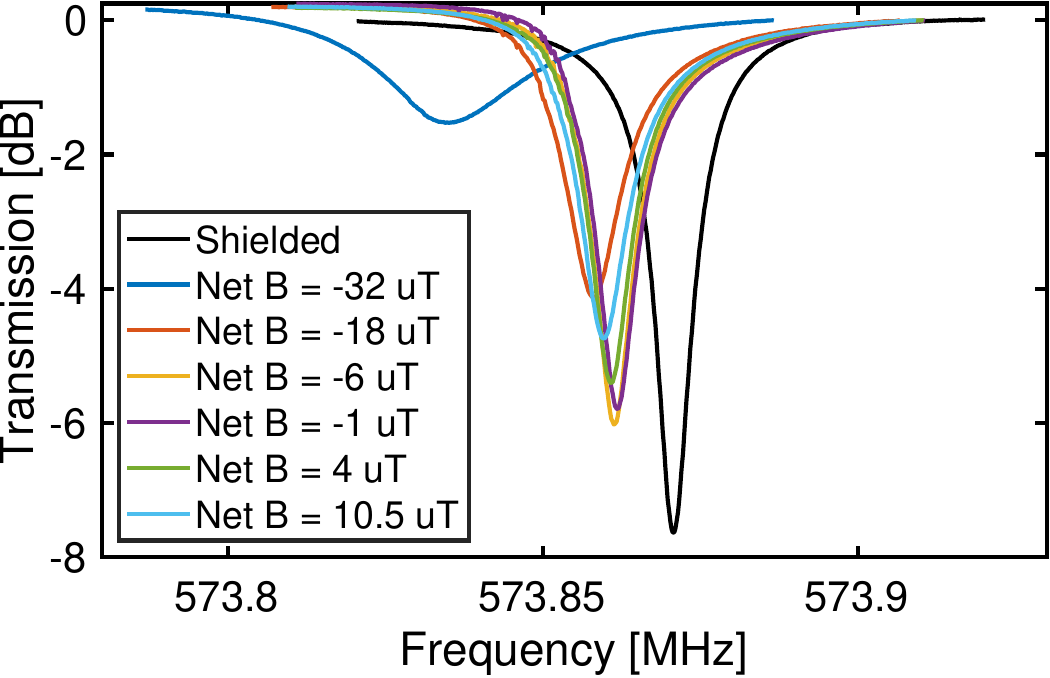}
   \end{tabular}
   \end{center}
   \caption[227] 
   { \label{fig:227} 
Measured normalized $|S_{21}|$ transmission profile as a function of frequency of a representative detector on the quadrant array. The transmission is shown for a variety of magnetic fields present when the aluminum housing and detectors cool through their $T_c$: The black line profile is a measurement when a mu-metal shield in present in the cryostat. The blue line represents a situation in which earth's field is present. Line profiles in other colors show the performance in different nulling fields. In these latter situations, the mu-metal shield is absent from the cryostat.}
   \end{figure} 
   
We also investigate the change of $S_{21}$ line profile of our KIDs, a representative is shown in Figure~\ref{fig:227}. The change in frequency and depth of the line profile indicates that as we tune the current flowing through the Helmholtz coil when we cycle the He-10 fridge, we are approaching on completely nulling the earth magnetic field perpendicular to the detector wafer. As the net absolute field is reduced the resonance frequency and the depth of transmission increase from the situation without magnetic shielding (blue line profile) to approaching the line profile when enclosed by the mu-metal magnetic shield (black line profile).

   \begin{figure} [ht]
   \begin{center}
   \begin{tabular}{c} 
   \includegraphics[height=7.5cm]{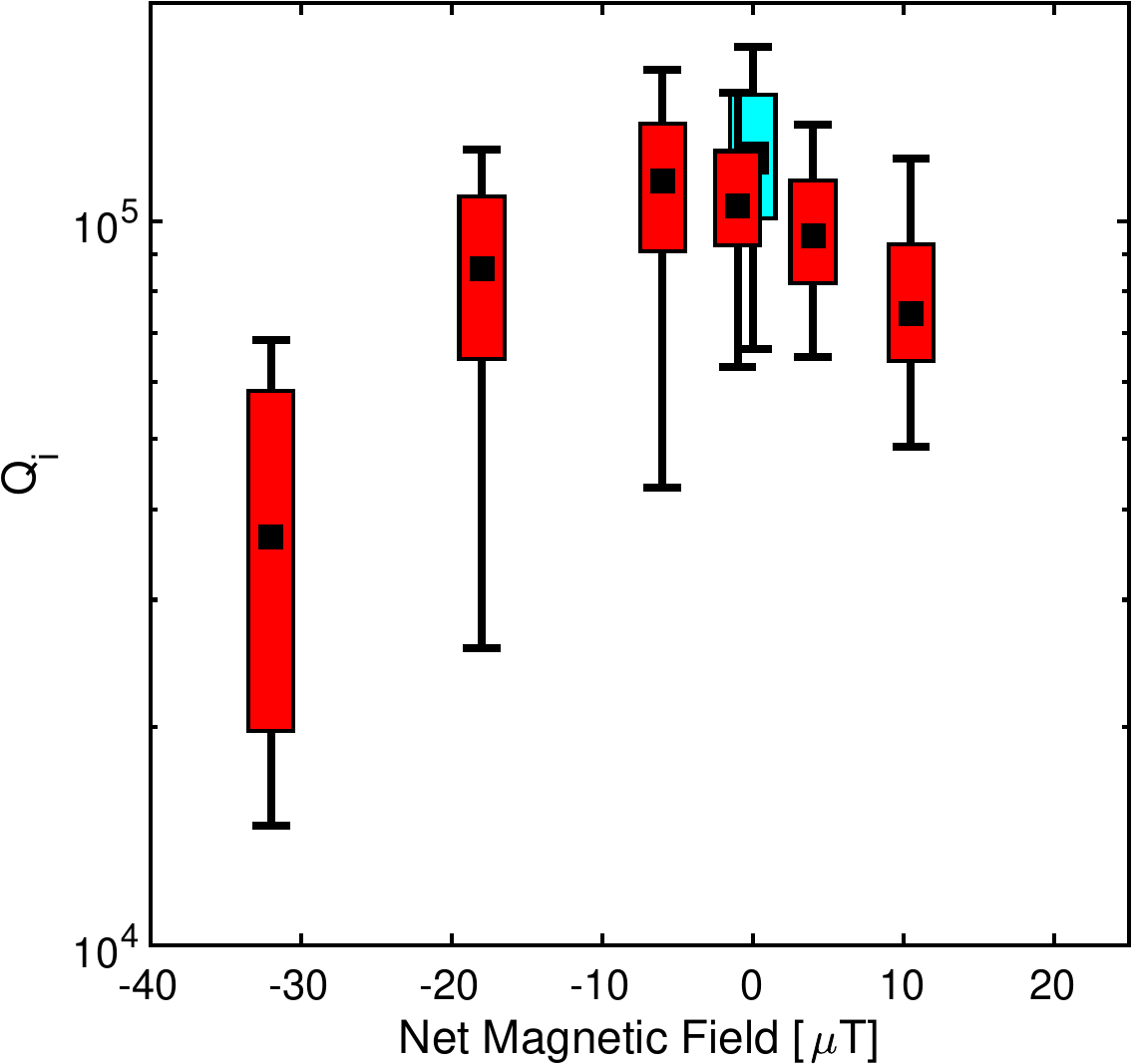}
   \end{tabular}
   \end{center}
   \caption[qi] 
   { \label{fig:qi} 
$Q_i$ vs. $B_{net}$, the net magnetic field through our detector wafer, where $B_{net} = B_{coil} - B_{earth}$ and $B_{earth} \sim 35$ $\mathrm{\mu T}$. Square markers are median values of $Q_i$; red bars are $1\sigma$ ranges, and error bars indicate $2\sigma$ ranges. As a comparison, we include the cyan plot for the $Q_i$ distribution with magnetic shielding in the cryostat.}
   \end{figure} 

We systematically study the $Q_i$ distribution under various nulling fields. Figure~\ref{fig:qi} and the bottom panel of Figure~\ref{fig:qfactor} show the results of this study. The trend of $Q_i$ shift in Figure~\ref{fig:qi} indicates that the $Q_i$ of most KIDs on our detector chip can be recovered maximally to $\sim 10^{5}$ when 0 $\mathrm{\mu T}<|B_{net}|<5$ $\mathrm{\mu T}$. Note that in our experiment it is difficult to completely remove the magnetic filed inside the cryostat, due to (1) the inaccuracy of the smartphone magnetometers and (2) the fact that the Helmholtz coil nulls the earth magnetic field perpendicular to the wafer but the angle difference exists between the Helmholtz coil and the total earth magnetic field. Based on the above studies, it is required to build a magnetic shield for the TIM full array, which can reach $|B| <$ 3 $\mathrm{\mu T}$ at the detector chips and which fits well in the TIM cryostat.

\section{SUMMARY AND PROSPECT}
\label{sec:summary}

We have designed and tested a quadrant-sized package for the KID arrays of TIM to demonstrate a novel approach to (1) maintain a 50 micron airgap accurately and reliably and (2) provide compliance to thermal deformations on cooldown. A VNA scan demonstrates $>90\%$ yield for our 864 pixel LW quadrant chip. Consistent resonator features over multiple thermal cycles demonstrate the robustness of our package design. We also study the magnetic field dependence of the quality factor of our KIDs and find that the earth magnetic field reduces the $Q_i$ by a factor of $2-5$, which causes higher loss within individual KIDs and degrades their performance. Based on our initial measurement results, we specify a requirement of $|B| <$ 3 $\mathrm{\mu T}$ for designing a magnetic shield to null the earth magnetic field inside the TIM cryostat.

\acknowledgments      
 
TIM is supported by NASA under grant 80NSSC19K1242, issued through the Science Mission Directorate. R.M.J. Janssen is supported by an appointment to the NASA Postdoctoral Program at the NASA Jet Propulsion Laboratory, administered by Oak Ridge Associated Universities under contract with NASA. Part of this research was carried out at the Jet Propulsion Laboratory, California Institute of Technology, under a contract with the National Aeronautics and Space Administration (80NM0018D0004).

\bibliography{report}
\bibliographystyle{spiebib}

\end{document}